\RequirePackage{fix-cm}
\documentclass{article}      
\usepackage[dvipdfmx]{graphicx}
\usepackage{amsmath,amssymb}
\begin{document}

\title{Multiplicative random cascades with additional stochastic process in
financial markets
}



\date{}
\maketitle

\centerline{ {\bf Jun-ichi Maskawa}$^{\rm a}$\footnote{\label{q}Jun-ichi Maskawa, 
Department of Economics, Seijo University
Address: 6-1-20 Seijo, Setagaya-ku, Tokyo 157-8511, Japan
Tel.: +81-3-3482-5938, Fax: +81-3-3482-3660
E-mail: maskawa@seijo.ac.jp }
 ,  {\bf Koji Kuroda}$^{\rm b}$ and {\bf Joshin Murai}$^{\rm c}$}

\vspace{10pt}

\centerline{$^{\rm a}$Department of Economics, Seijo University} \vspace{10pt}

\centerline{$^{\rm b}$Graduate School of Integrated Basic Sciences, Nihon University}

\vspace{10pt}

\centerline{$^{\rm c}$Graduate School of Humanities and Social Sciences, Okayama University}
\vspace{10pt} \vspace{10pt}

\begin{abstract}
Multiplicative random cascade model naturally reproduces the intermittency or multifractality, which is frequently shown among hierarchical complex systems such as turbulence and financial markets. As described herein, we investigate the validity of a multiplicative hierarchical random cascade model through an empirical study using financial data. Although the intermittency and multifractality of the time series are verified, random multiplicative factors linking successive hierarchical layers show strongly negative correlation. We extend the multiplicative model to incorporate an additional stochastic term. Results show that the proposed model is consistent with all the empirical results presented here.
\end{abstract}

\section{Introduction}
\label{intro}

Financial markets consist of different participants who watch the market with different temporal resolution and who react to price changes with different time horizons. For example, whereas intra-day traders watch the market continuously and trade multiple times each day, fund managers might reconsider their portfolios over periods of weeks or months. Competition among participants with different characteristic time horizons creates a heterogeneous structure of volatilities measured using different time resolutions. Past coarse-grained measures of volatility correlate to future fine-scale volatility more than the reverse process. This causal structure from a long-term to short-term scale volatility was first pointed out by M\"uller et al. (M\"uller et al. 1997). It is thought to be a stylized fact in financial markets (Arneodo et al. 1998a; Cont 2001; Lynch and Zumbach 2003), leading naturally to the idea of volatility cascade from long-term to short-term horizon by an analogy of an energy cascade in turbulence in fluid dynamics (Ghashghaie 1996). In fully developed turbulent flow, the kinetic energy is injected by external forces to create eddies at the largest space scale. According to the phenomenological picture of hierarchical cascade in turbulence, they are deformed by fluid dynamics, with breakage into smaller eddies. Then the energy transfers to a smaller scale. This process is repeated hierarchically several times to the smallest scale, where the energy is eventually removed by dissipation (Richardson 1922; Kolmogorov 1941; Frisch 1997).  

Intermittency, a common phenomenon in many complex systems, is also a crucial characteristic both in financial markets and turbulent flows (Kolmogorov 1962; Mandelbrot 1963). Intermittency is characterized by the presence of irregular burst of the volatility of asset prices, the strength of velocity of fluids and other quantities of systems and creation of large kurtosis and fat tails in probability distribution of those quantities. One also frequently observes multifractality in the same systems showing intermittency (Frisch 1997; Schmitt et al. 1999). Financial markets and turbulence are not exceptions. Multiplicative random cascade models that will be introduced into section 2 relate the fluctuation at scale $\lambda l$ ($\lambda<1$) to that at scale $l$ by the cascade rule
\begin{equation}
\delta_{\lambda l} X(t)=W_{\lambda} \delta_l X(t),
\label{cascade}
\end{equation}
where $W_{\lambda}$ is a random variable that depends only on the scale ratio $\lambda$.
It is a promising model leading naturally to the intermittency and multifractality of the systems (Mandelbrot 1974; Frisch 1997; Arneodo et al. 1998b)

As described in this paper, we investigate the validity of a multiplicative hierarchical random cascade model by application of the model to a time series of stock prices. Because the actual hierarchical structure is directly unobservable, we apply a model on a dyadic hierarchical tree structure to the market. The intermittency and multifractality of the time series are verified with prediction of the model. However, although random multiplicative factors linking successive hierarchical layers are assumed to be i.i.d., the corresponding values calculated backwards from the data show strongly negative correlation. That result apparently indicates a need for extension of the model. In each cascading step (\ref{cascade}) the variable $\delta_{\lambda l} X(t)$ is usually determined by only one predecessor $\delta_l X(t)$ at the previous step. We extend the multiplicative model to incorporate an additional stochastic term multiplied by the standard deviation of the variable at the previous step. Such a model was introduced originally by Jim\'enez as a mixed multiplicative-stochastic model of turbulence (Jim\'enez 2000; Jim\'enez 2007). Our model is shown to cope with both well-known characteristics of financial time series such as intermittency or multifractality and the observed negative correlation among multiplicative factors.

The paper is organized as follows. In Section \ref{sec:2}, we introduce a multiplicative random cascade model on wavelet dyadic trees proposed by Arneodo, Bacry and Muzy (Arneodo et al. 1998b), which is the model investigated herein. We next investigate the model validity by application of the model to a time series constructed from the multivariate time series of the stock prices of the constituents of the FTSE 100 listed on the London Stock Exchange. In Section \ref{sec:4}, we introduce an extension of the model and verify the improved model validity by Monte Carlo simulations. Finally, we summarize the results and provide some remarks about future work.

\section{Multiplicative dyadic random cascade model}
\label{sec:2}
Arneodo, Bacry and Muzy proposed a method of constructing a new class of random functions, designated as $\mathcal{W}$-cascade, using the orthogonal wavelet transform (Arneodo et al. 1998b). We briefly introduce $\mathcal{W}$-cascade in this section.  

Function $f(x) \in L^2(\mathbb{R})$ can be expanded by an orthogonal wavelet basis $\{\phi_{j_0,k_(x)},\psi_{j,k_(x)}\}_{j_0 \le j,k\in\mathcal{Z}}$ as
\begin{equation}
f(x)=\textstyle\sum\limits_{k\in\mathcal{Z}}c_{j_0,k}\phi_{j_0,k}(x)+\textstyle\sum\limits_{j=j_0}^\infty\textstyle\sum\limits_{k\in\mathcal{Z}}d_{j,k}\psi_{j,k}(x),
\label{we} 
\end{equation}
where the integer $j_0$ can be arbitrarily chosen.
A wavelet basis, functions $\phi_{j,k}(x)$ and $\psi_{j,k}(x)$, can be constructed respectively using translation and dilation of the scaling function $\phi(x)$ and the wavelet function $\psi(x)$:
\begin{equation}
\phi_{j,k}(x)=2^{j/2}\phi(2^{j}x-k),~\psi_{j,k}(x)=2^{j/2}\psi(2^{j}x-k),~0 \le j~,0 \le k < 2^{j-1}.
\label{wb}
\end{equation}
The wavelet coefficients $\{c_{j_0,k},d_{j,k}\}$ of a function $f(x)$ are then given by the integration shown below.
\begin{equation}
c_{j,k}=\int_{-\infty}^{\infty}f(x)\phi_{j,k}(x)dx,~d_{j,k}=\int_{-\infty}^{\infty}f(x)\psi_{j,k}(x)dx
\label{wc}       
\end{equation}
In the following, we specifically address discrete time series with finite length $L$. When applying equations (\ref{we})--(\ref{wc}), the function $f(x)$ is used as a periodic function of length $L$. The largest scale $s_0=L$ corresponding to $j=0$ and the finer scales $s_j=L/2^j~(j=1,\dots,\log_2(L))$. 
 The wavelet function is expected to have $N_{\psi}~( \ge 1)$ vanishing moments
\begin{equation}
\int_{-\infty}^{\infty}x^n\psi(x)dx=0,~0 \le n < N_{\psi}.
\label{vm}
\end{equation}

\begin{figure*}
  \includegraphics[width=0.75\textwidth]{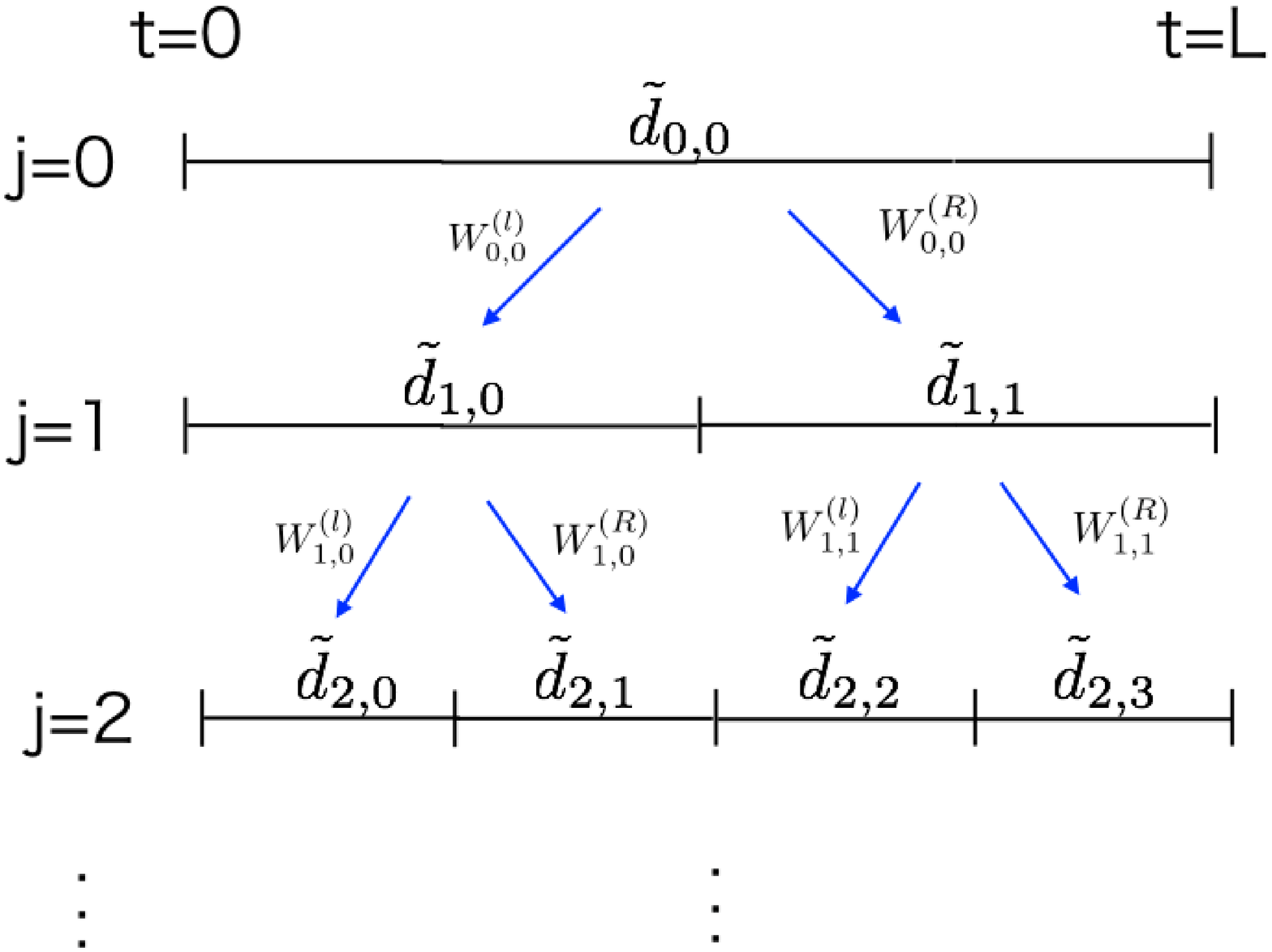}
\caption{Schematic draw of $\mathcal{W}$-cascade.}
\label{fig:cascade}       
\end{figure*}

Arneodo et al. have built a random function $f(x)$ by specifying the wavelet coefficients $\{c_{j_0,k},d_{j,k}\}_{j_0 \le j,k\in\mathcal{Z}}$. We set $j_0=0$. The coefficients $c_{0,0}$ and $d_{0,0}$ are chosen as arbitrary numbers\footnote{When we apply the model to the time series as in section 3, the values $c_{0,0}$ and $d_{0,0}$ are fixed by the equation (\ref{wc}).}. Rescaling of the coefficients as $\tilde{d}_{i,k}=2^{j/2}d_{i,k}$ is useful for simplifying the equation. Coefficients $\tilde{d}_{i,k}$ are defined recursively as
\begin{equation}
\tilde{d}_{j,2k}=W^{(l)}_{j-1,k}~\tilde{d}_{j-1,k},~\tilde{d}_{j,2k+1}=W^{(r)}_{j-1,k}~\tilde{d}_{j-1,k},~1 \le j~,0 \le k < 2^{j-1},
\label{w}
\end{equation}
where the multiplicative factors $\{W^{(\epsilon)}_{j-1,k}\}_{k\in\mathcal{Z},\epsilon=l,r}$ are independent identically distributed (i.i.d.) real-valued random variables. The $\mathcal{W}$-cascade is presented schematically in Fig. \ref{fig:cascade}. First, length $L$ is divided into two intervals. The wavelet coefficients $\tilde{d}_{1,0}$, and $\tilde{d}_{1,1}$ are defined by the multiplicative factors $W^{(l)}_{0,0}$ and $W^{(r)}_{0,0}$. Next, each divided interval is divided into two intervals again. The wavelet coefficients $\tilde{d}_{2,k}$ ($k=0,\dots,3$) on those intervals are defined by multiplicative factors $W^{(\epsilon)}_{1,k}$ ($\epsilon=l,r,~k=0,1$). This process on the dyadic tree is repeated to the finest scale $s_j=2$.

For $\mathcal{W}$-cascade, it is easy to derive the self-similarity law of the probability density function (PDF) of wavelet coefficients. In fact,
the PDF $P_j(\tilde{d})$ of wavelet coefficients $\tilde{d}_{j,k}=2^{j/2}d_{j,k}$ derived by the equation (\ref{wc}) at a scale $s_j$ are linked to the PDF $P_{j'}(\tilde{d})$ at another scale $s_{j'}$ through the self-similar kernel $G_{jj'}(x)$
\begin{equation}
P_j(\tilde{d})=\int G_{jj'}(x)e^{-x}P_{j'}(e^{-x}\tilde{d})dx,
\label{ss}
\end{equation}
where the self-similar kernel $G_{jj'}(x)$ is the PDF of logarithm of the product $x=\log(W_{j'}W_{j'+1},\dots,W_{j-1})=\sum_{k=j'}^{j-1} \log(W_k)$.
When the path of the process $f(x)$ is fractal (homogeneously scale invariant) with the H\"{o}lder exponent $H$, it satisfies
\begin{equation}
P_j(\tilde{d})=(\frac{s_j}{s_{j'}})^{-H}P_{j'}((\frac{s_j}{s_{j'}})^{-H}\tilde{d}).
\label{ssa}
\end{equation}
They have also derived the singular-spectrum, and auto-correlation functions analytically for $\mathcal{W}$-cascade. Their functions possess properties covering a large part of well known stylized facts of financial time series.  

\section{Empirical study of $\mathcal{W}$-cascade}
\label{sec:3}

We investigate the validity of a multiplicative hierarchical random cascade model by application of the model to a time series of stock price. Because the actual hierarchical structure is not directly observable, we apply a model on a dyadic hierarchical tree structure to the market, i.e., $\mathcal{W}$-cascade. 

We analyze the normalized average of the logarithmic stock prices of the constituent issues of FTSE 100 index listed on the London Stock Exchange for the period from November 2007 to January 2009, which includes the Lehman shock on 15 September 2008 and the market crash on 8 October 2008.

\begin{figure*}
  \includegraphics[width=0.75\textwidth]{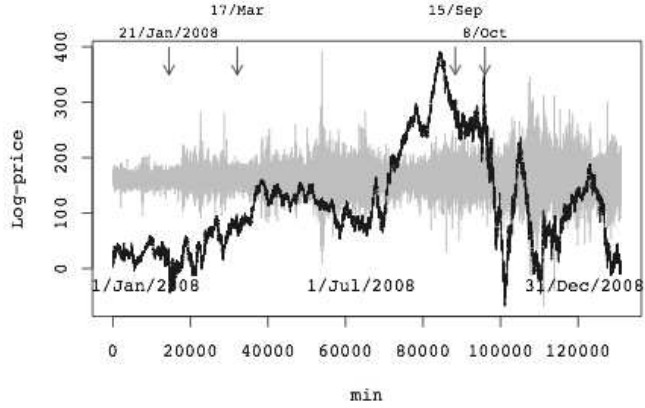}
\caption{Path of the process (black line) constructed by the accumulation of 1 min. log-returns in the daytime (grey line). Total:$2^{17}=131,072$ min. See text for details.}
\label{fig:path}       
\end{figure*}

First, we calculate the average of deseasonalized return of each issue $\delta X_i(t)=\log(P_i(t))-\log(P_i(t-\delta t))$
\begin{equation}
\delta f(k\delta t)=\frac{1}{N_F}\sum_{i=1}^{N_F}\frac{\delta X_i(k\delta t)-\mu_i}{\sigma_i},
\label{df}
\end{equation}
where $\mu_i$ and $\sigma_i$ respectively denote the average and the standard deviation of $\delta X_i$ and $N_F$ is the number of constituent issues. Here, we set $\delta t=1$ and examine the 1-min. log-return.
We excluded the overnight price change and specifically examine the intraday evolutions of returns. To remove the effect of intraday U shape pattern of market activity from the time-series, the return was divided by the standard deviation of the corresponding time of the day for each issue $i$.
Then we cumulate $\delta f(t)$ to obtain the path of the process  $f(k\delta t)~(k=1,\dots,L)$ (Fig. \ref{fig:path}) as
\begin{equation}
f(k\delta t)=\sum_{k'=1}^k \delta f(k'\delta t).
\label{f}
\end{equation}

%

\begin{figure*}
  \includegraphics[width=0.75\textwidth]{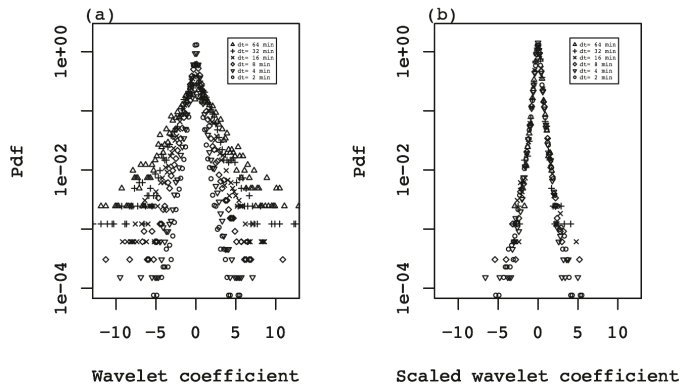}
\caption{Probability density function of wavelet coefficients. (a) PDF of $\tilde{d}_{j,k}$. (b) PDF of scaled coefficients $\tilde{d}_{j,k}/(s_j)^H$. We chose the value $H$=0.53 taking the peak of the singular spectrum explained below. See also Fig. \ref{fig:multifractal}(c). }
\label{fig:ss}       
\end{figure*}

We use the Daubechies 4 compactly supported orthogonal wavelet basis (Daubechies 1992) for the following, which has $N_{\psi}=2$ vanishing moments.
In Fig. \ref{fig:ss}(a), we present the PDF of $\tilde{d}_{j,k}$ for the scales $s_j=2,4,\dots,64$ min. The tails of the PDF become fatter as the corresponding scale $s_i$ becomes larger. However, the PDFs of scaled wavelet coefficients $\tilde{d}_{j,k}/(s_j)^H$ for different scales collapse into a single curve as shown in Fig. \ref{fig:ss}(b) when we set the parameter $H$ near 0.5. As described by the equation (\ref{ssa}), the result indicates that Brownian motion approximates the process well.

Next, we analyze the multifractal properties of the path $f(x)$ by an approach using a wavelet-based multifractal formalism proposed by Muzy, Bacry, and Arneodo (Muzy et al. 1993). At the beginning, we define two mathematical terms. The H\"{o}lder exponent $\alpha(x_0)$ of a function $f(x)$ at $x_0$ is defined as the largest exponent such that there exists an nth order polynomial $P_n(x)$ and constant $C$ that satisfy
\begin{equation}
|f(x)-P_n(x-x_0)| \le C|x-x_0|^{\alpha}
\label{holder},
\end{equation}
for $x$ in a neighborhood of $x_0$, characterizing the regularity of the function $f(x)$ at $x_0$. The singular spectrum $D(\alpha)$ is the Hausdorff dimension of the set where the H\"{o}lder exponent is equal to $\alpha$,
\begin{equation}
D(\alpha)=dim_H\{x|\alpha(x)=\alpha\}.
\label{singular}
\end{equation}
For multifractal paths, the H\"{o}lder exponent $\alpha$ distribute in a range, while for paths of the Brownian motion, which is fractal, $D(0.5)=1$ and $D(\alpha)=0$ for $\alpha\ne0.5$.

\begin{figure*}
  \includegraphics[width=0.75\textwidth]{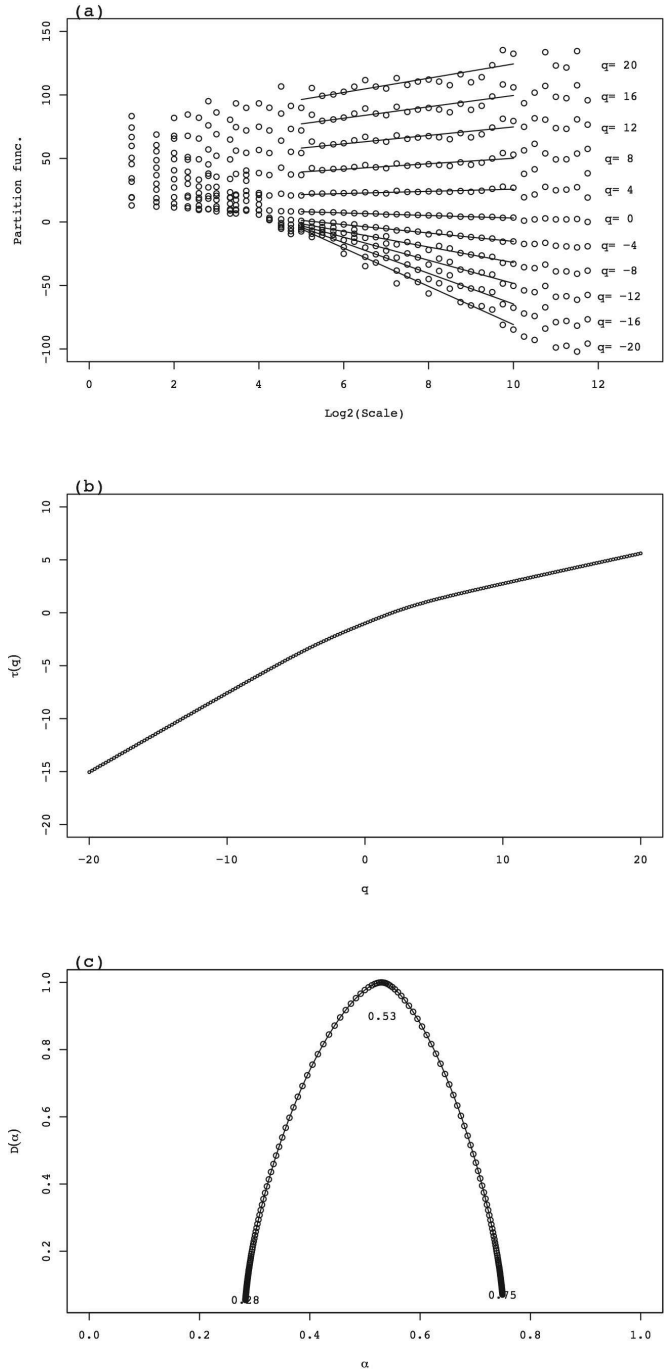}
\caption{Multifractal analysis of the data. (a) Log--log plots of partition functions $Z(q,s)$ vs. scale $s$ for the orders $q=-20,\dots,20$ ($\circ$) and their least square linear fits (solid lines). (b) Exponent $\tau(q)$ as a function of $q$. (c) Singular spectrum $D(\alpha)$.}
\label{fig:multifractal}       
\end{figure*}

Muzy, Bacry and Arneodo proposed the wavelet transform modulus maxima (WTMM) method based on continuous wavelet transform of function to calculate the singular spectrum $D(\alpha)$ (Muzy et al. 1993). We briefly sketch the WTMM method in Appendix. We calculate the partition function $Z(q,s)$ of the $q$-th moment of wavelet coefficients by the equation (\ref{z}) for the path of our data. Results are shown in Fig. \ref{fig:multifractal}(a). The partition function $Z(q,s)$ for each order $q$ shows power law behavior in the range of scales $<2^{10-11}$. Exponents $\tau(q)$ are derived by the equation (\ref{tau}). As shown in Fig. \ref{fig:multifractal}(b), it is a convex function of $q$. Those results show the multifractality of the path of the data. The singular spectrum $D(\alpha)$ derived, as the Legendre transformation of the function $\tau(q)$, by the equation (\ref{D}) is a convex function that has compact support $[0.28,0.75]$ taking the peak at $\alpha=0.53$, as shown in Fig. \ref{fig:multifractal}(c).

\begin{figure*}
  \includegraphics[width=0.75\textwidth]{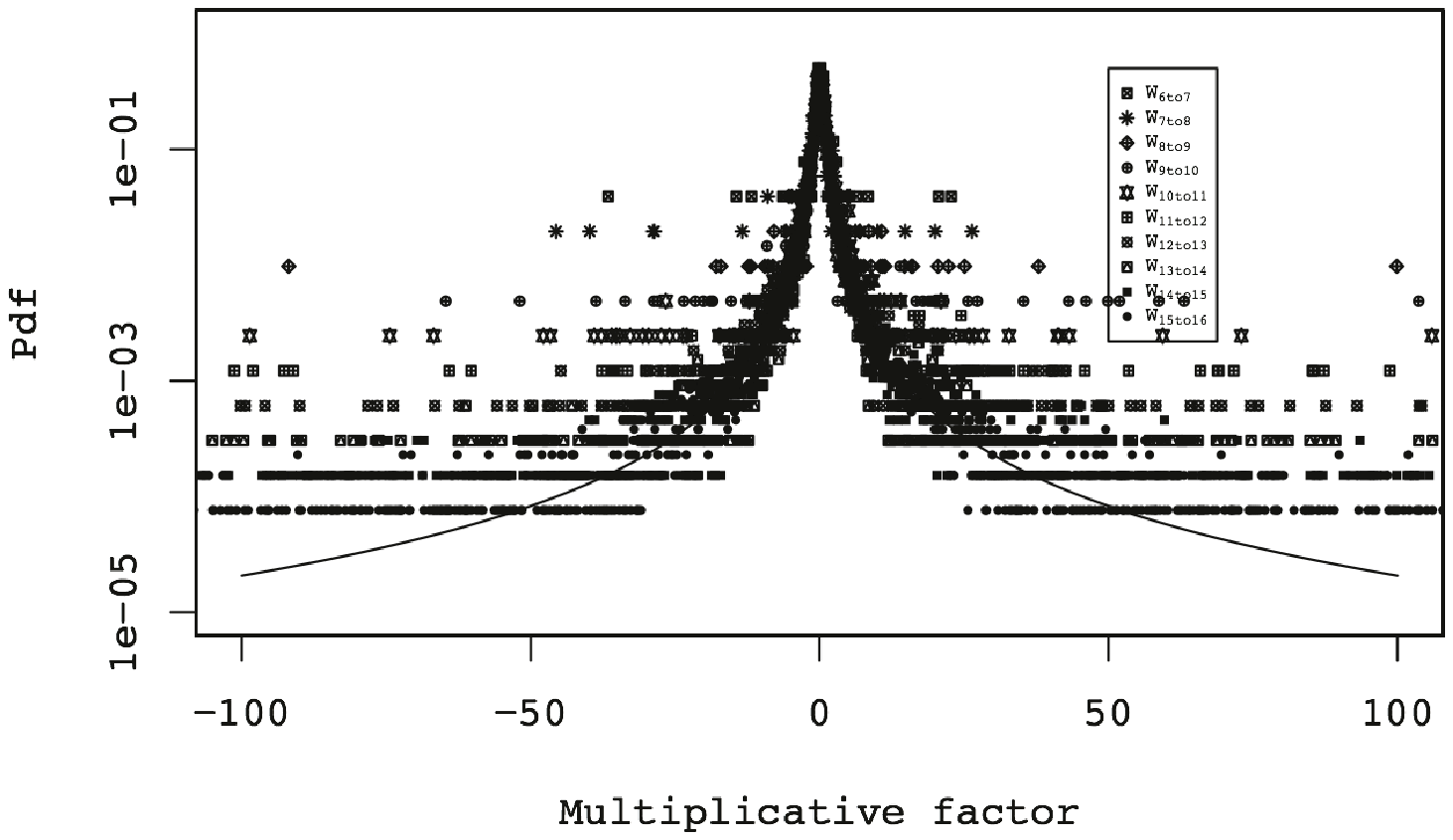}
\caption{The PDFs of multiplicative factors calculated from the wavelet coefficients of the data. The curve is the least squares fit to the Cauchy's PDF $p(x)=(\pi s(1+\frac{x^2}{s^2}))^{-1}$ ($s=0.6$).}
\label{fig:pdf-w}       
\end{figure*}

\begin{figure*}
  \includegraphics[width=0.75\textwidth]{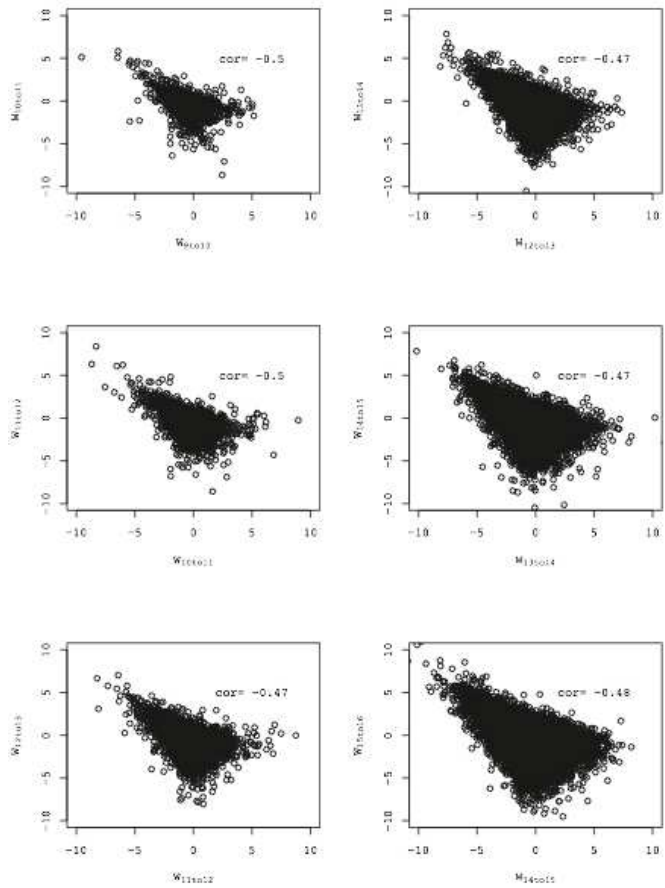}
\caption{Scatter plots of successive multiplicative factors $W_{j-1toj}$ and $W_{jtoj+1}$ for $j=10,\dots,15$. The sample correlation coefficient between two qualities is shown in each panel.}
\label{fig:w-w}       
\end{figure*}

\begin{figure*}
  \includegraphics[width=0.75\textwidth]{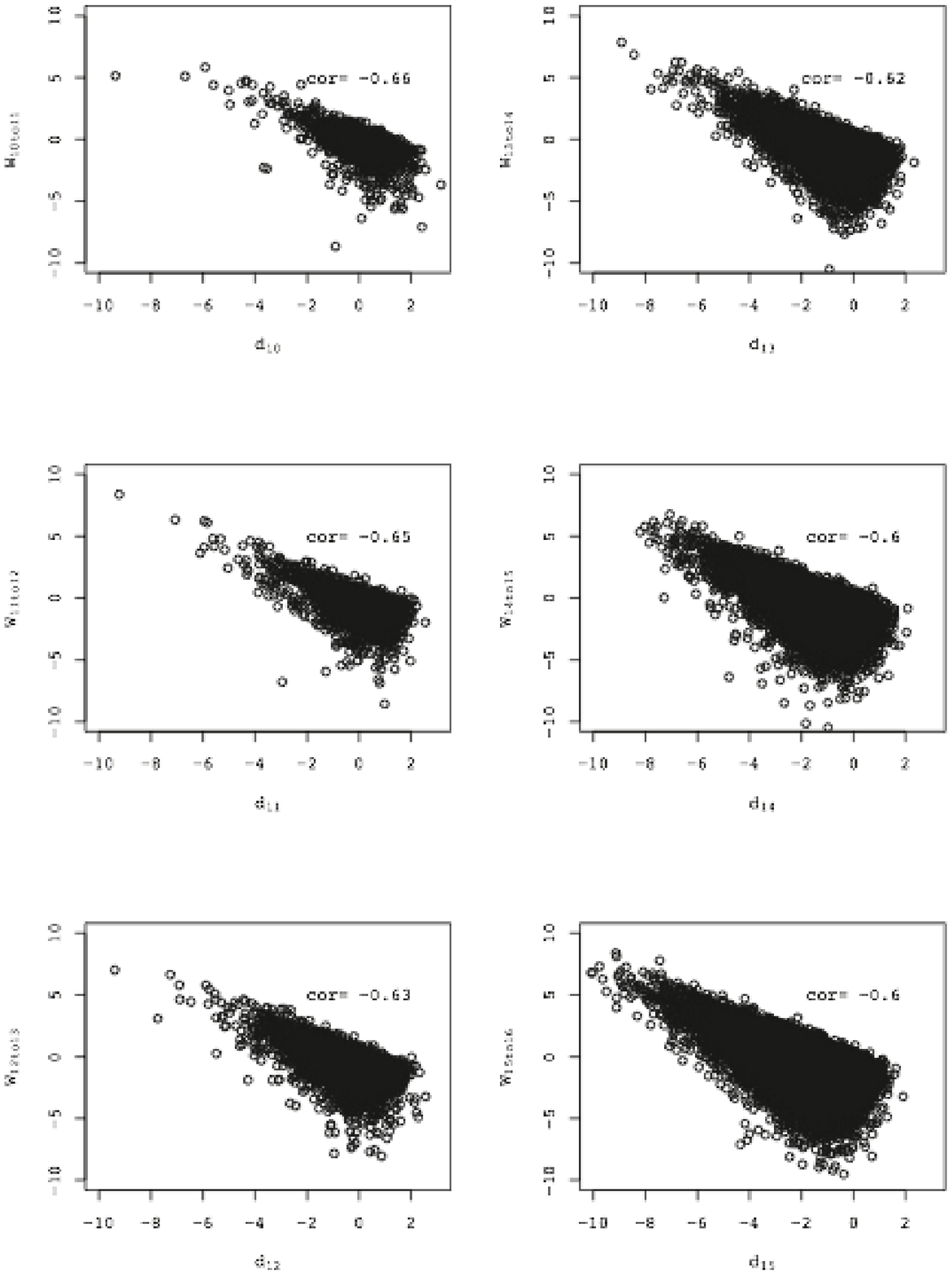}
\caption{Scatter plots of predecessor $\tilde{d}_j.$ and multiplicative factor $W_{jtoj+1}$ for $j=10,\dots,15$. The sample correlation coefficient between two qualities is shown in each panel.}
\label{fig:d-w}       
\end{figure*}

In  $\mathcal{W}$-cascade model random multiplicative factors $W^{\epsilon}_{j,k}$ linking successive hierarchical layers j and j+1 are assumed to be i.i.d. We calculate the multiplicative factors $W^{\epsilon}_{j,k}$ backward from wavelet coefficients of the data and compile statistics of those qualities. 
Fig. \ref{fig:pdf-w} shows the PDFs of multiplicative factors $W^{l}_{j,k}$ calculated from the wavelet coefficients of the data\footnote{The statistics of the multiplicative factors $W^{r}_{j,k}$ is the same as of $W^{l}_{j,k}$. It is not shown here.} . The multiplicative factors are broadly distributed and are well fitted by the Cauchy's PDF irrelevant to the layer $i$. An important difference between the data and model is the strongly negative correlation between successive multiplicative factors and predecessor wavelet coefficients and multiplicative factors. Results are shown in Fig. \ref{fig:w-w} and Fig. \ref{fig:d-w}.


\section{Multiplicative cascade with additional stochastic process}
\label{sec:4}
In $\mathcal{W}$-cascade model (\ref{w}), in each cascading step, the wavelet coefficients $\tilde{d}_{j+1,2k}$ and $\tilde{d}_{j+1,2k+1}$ locally transit from the only one predecessor $\tilde{d}_{j,k}$. We extend the multiplicative model to incorporate a certain range of predecessors as an additional stochastic term. Such extension is introduced originally by Jim\'enez as a mixed multiplicative-stochastic model of turbulence (Jim\'enez 2000; Jim\'enez 2007). He has proposed a model incorporating the global standard deviation of the velocity increment of fluid describing a trigger of cascade by the properties of the surrounding fluid.

\begin{figure*}
  \includegraphics[width=0.75\textwidth]{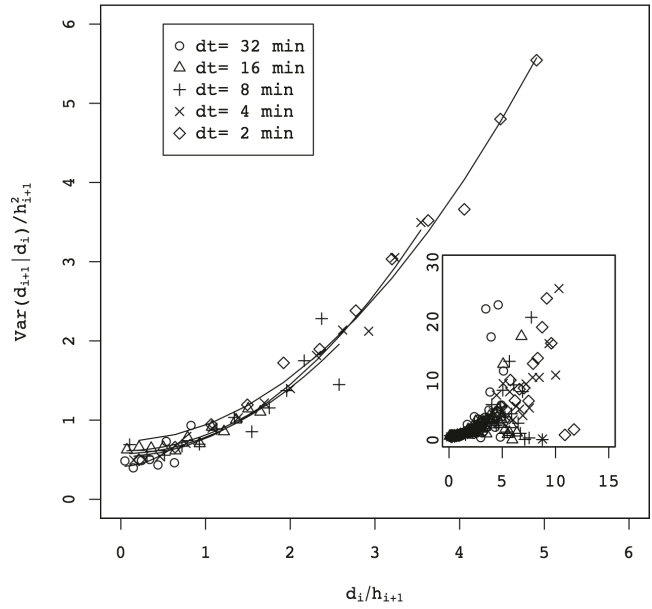}
\caption{Conditional variance of successor $\tilde{d}_{j+1.}/h_{j+1}$ as a function of predecessor $\tilde{d}_{j.}/h_{j+1}$. Data are divided into bins with the same intervals 0.2. Bins with fewer than 100 data are excluded from the main panel, while the inset includes all bins.}
\label{fig:8}       
\end{figure*}

As a first step along the direction, we investigate the multiplicative cascade model with an additional stochastic term proportional to the sample standard deviation $h_j=Std(\tilde{d}_j.)$ as
\begin{equation}
\tilde{d}_{j+1,2k}=W^{(r)}_{j,k}\tilde{d}_{j,k}+\eta^{(r)}_{j,k}h_j,~\tilde{d}_{j+1,2k+1}=W^{(l)}_{j,k}\tilde{d}_{j,k}+\eta^{(l)}_{j,k}h_j
\label{model}
\end{equation}
where both multiplicative and additional stochastic variables $\{W^{(l/r)}_{j,k}\}_{j,k}$ and $\{\eta^{(l/r)}_{j,k}\}_{j,k}$ are assumed to be i.i.d. with zero mean and they are independent.
Obtaining the conditional and non-conditional variance of both sides of the equation (\ref{model}), we have two equations
\begin{equation}
Var(\tilde{d}_{j+1.}|\tilde{d}_{j.})=Var(W^{(l/r)}_{j.})\tilde{d}_{j.}^2+Var(\eta^{(l/r)}_{j.})h_j^2
\label{var1}
\end{equation}
and
\begin{equation}
(\frac{h_{j+1}}{h_j})^2=Var(W^{(l/r)}_{j.})+Var(\eta^{(l/r)}_{j.}).
\label{var2}
\end{equation}
Equation (\ref{var1}) is investigated using the data presented in the previous section. The results for several time scales are shown in Fig. \ref{fig:8}. We perform linear regression analysis $Y=aX+b$ where the explained variable $Y$ is the conditional variance of successor divided by the non-conditional variance $Var(\tilde{d}_{j+1.}|\tilde{d}_{j.})/h_{j+1}^2$. The explanatory variable $X$ is the square of predecessor $\tilde{d}_{j.}^2/h_{j+1}^2$.
The results are shown in Table \ref{tab:1}. The regression lines are added to Fig. \ref{fig:8}. The mean values of the variance $Var(W)$ and $Var(\eta) $ respectively denote 0.18 (0.08) and 0.32 (0.07), in which the standard deviations are presented in brackets. The results are consistent with i.i.d. assumption for the stochastic variables $W^{(l/r)}_{j,k}$ and $\eta^{(l/r)}_{j,k}$ within the standard deviations, except for two extreme samples 12(r) and 16(l).
Similar results have been obtained for turbulence experiments (Jim\'enez 2007).

\begin{table}
\caption{Results of regression analysis $Y=aX+b$. The variance $Var(W^{(l/r)}_{j.})$ and $Var(\eta^{(l/r)}_{j.})$ are derived from equations (\ref{var1}) and (\ref{var2}). }
\label{tab:1}       
\begin{tabular}{llllllll}
\hline\noalign{\smallskip}
Scale i (left/right) & a & b & Std of a & Std of b & Adj. $R^2$ & Var(W) & Var($\eta$) \\
\noalign{\smallskip}\hline\noalign{\smallskip}
12(l)	&0.61	&0.10	&0.05	&0.12	&-0.04	&0.19	&0.30\\
12(r)	& 0.42	& 0.63	& 0.06	& 0.19	& 0.57	& 0.35	& 0.25\\
13(l)	&0.66	&0.25	&0.13	&0.09	&0.36    & 0.17	&0.33\\
13(r)	& 0.61	& 0.20	& 0.03	& 0.02	& 0.91	& 0.20	& 0.31\\
14(l)	&0.72	&0.13	&0.04	&0.01	&0.88   &0.13	&0.34\\
14(r)	& 0.57	& 0.21	& 0.10	& 0.03	& 0.77	& 0.21	& 0.28\\
15(l)	&0.68	&0.17	&0.04	&0.01	&0.98	&0.14	&0.31\\
15(r)	& 0.54	& 0.23	& 0.06	& 0.01	& 0.98	& 0.21	& 0.25\\
16(l)	&0.90	&0.15	&0.15	&0.01	&0.92	&0.05	&0.47\\
16(r)	& 0.73	& 0.20	& 0.08	& 0.01	& 0.99	& 0.14	& 0.38\\
\noalign{\smallskip}\hline
\end{tabular}
\end{table}

\begin{figure*}
 \includegraphics[width=0.75\textwidth]{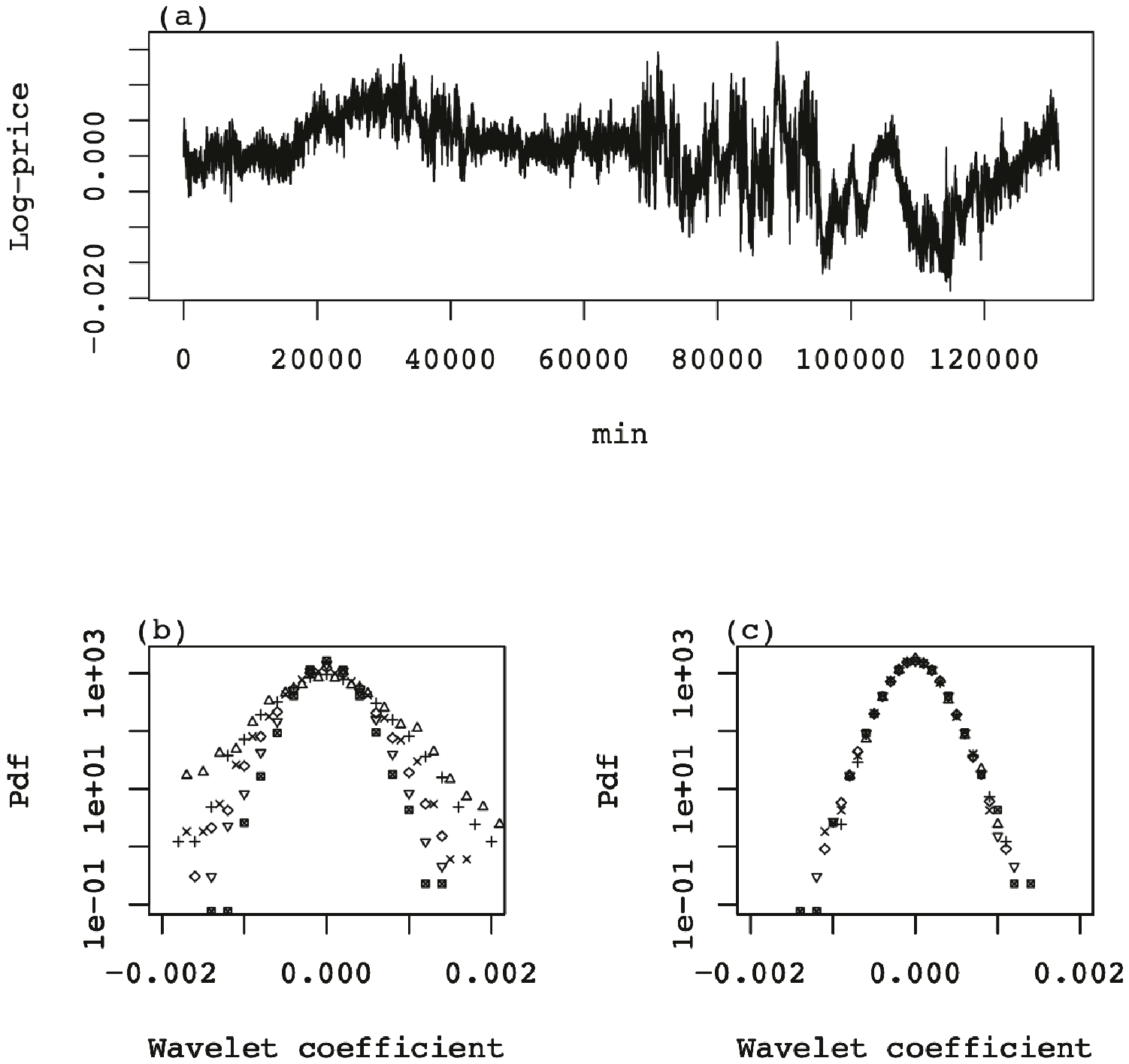}
\caption{Path of multiplicative cascade with additional stochastic process. (a) Path is reconstructed from the wavelet coefficients created from equation (\ref{model}). We use the Daubechies four compactly supported orthogonal wavelet basis. The stochastic variables $\log|W|\sim N(-0.33\log2,0.02\log2)$ and $\eta^{(l/r)}_{j,k}\sim N(0,0.3)$. (b) The PDF of the created wavelet coefficients $\tilde{d}_{j,k}$. (c) The PDF of scaled coefficients $\tilde{d}_{j,k}/(s_i)^H$. We choose the H\"{o}lder exponent $H$=0.23 taking the peak of the singular spectrum explained below. See also Fig. \ref{fig:10}. }
\label{fig:9}       
\end{figure*}

\begin{figure*}
  \includegraphics[width=0.75\textwidth]{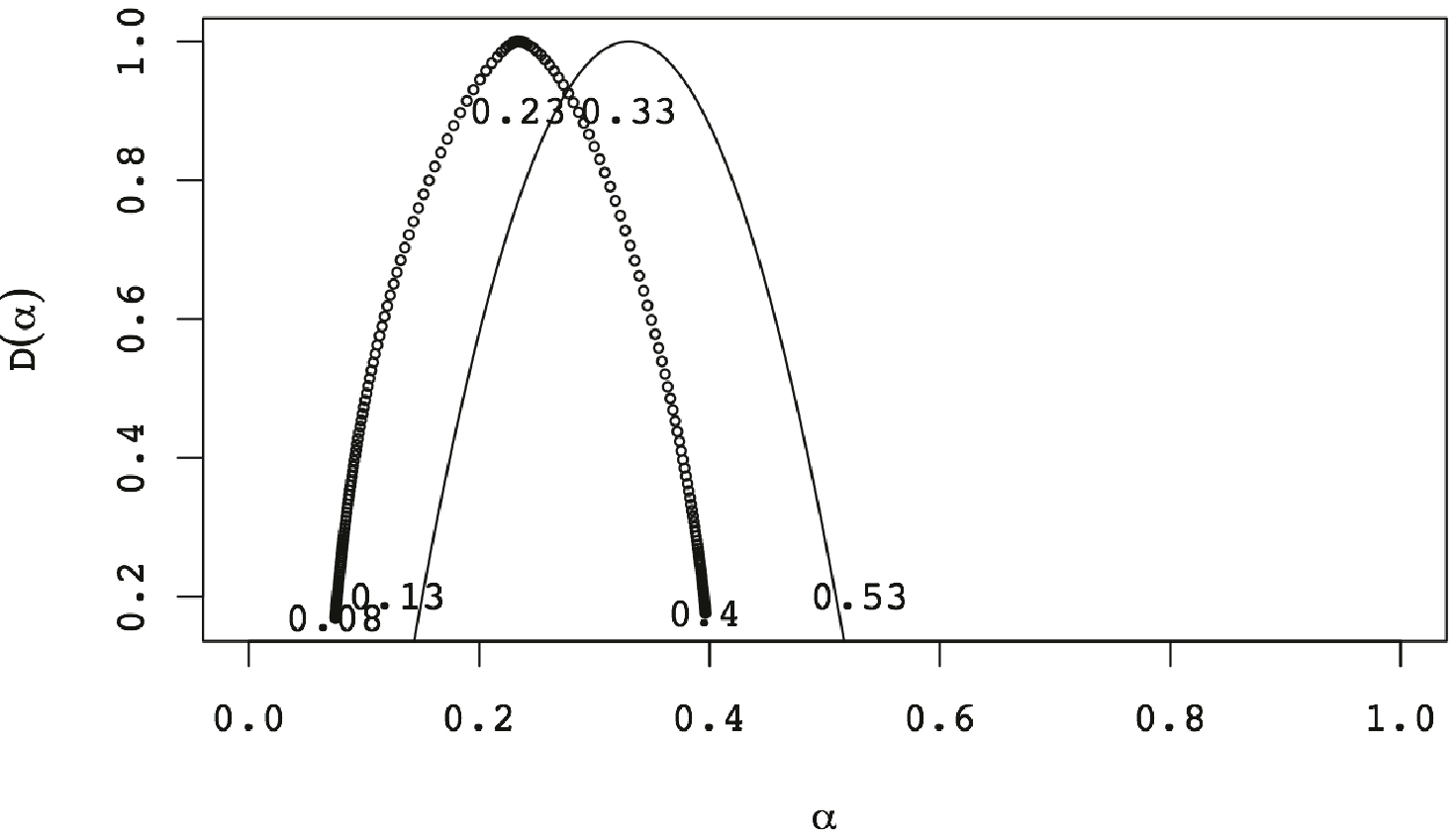}
\caption{Singular spectrum $D(\alpha)$ ($\circ$) of the realization Fig. \ref{fig:9}(a). Solid line shows the theoretical singular spectrum of the path of the multiplicative process with the stochastic variables $\log|W|\sim N(-0.33\log2,0.02\log2)$ (Arneodo et al. 1998b).}
\label{fig:10}       
\end{figure*}

In Fig. \ref{fig:9}, we present a realization of multiplicative cascade with additional stochastic process (\ref{model}) in which the stochastic variables $W^{(l/r)}_{j,k}$ and $\eta^{(l/r)}_{j,k} $ respectively denote drawn from the folded lognormal distribution and normal distribution. As presented in Fig. \ref{fig:9}(c), the PDFs of scaled wavelet coefficients collapse into a curve representing the self-similarity (\ref{ssa}) with the H\"{o}lder exponent $H=0.23$, which indicates that the fractional Brownian motion approximates the process well. The singular spectrum $D(\alpha)$ is a convex function that has the compact support $[0.08,0.40]$ taking the peak at $\alpha=0.23$, which is presented in Fig. \ref{fig:10}. The multifractality of the path is not broken by the additional term in the realization. 


Finally, we assess the data of the ratio $\tilde{d}_{j+1.}/\tilde{d}_{i.}$, which corresponds to $W^{(l/r)}_{j,k}$ of multiplicative cascade models. The PDF of the ratio still has a broad distribution and is well fitted by the Student's t distribution with 2 degrees of freedom irrelevant to layer $j$. More importantly, the strongly negative correlation between successive multiplicative factors and predecessor wavelet coefficients and multiplicative factors are reproduced as in the actual data\footnote{The results are not shown here.}.

\section{Conclusions}
\label{sec:5}
We investigated the validity of multiplicative random cascade model through an empirical study of the time series of the averaged day-time stock prices of the constituents of the FTSE 100 Index listed on the London Stock Exchange during Nov. 2007 -- Jan. 2009. The intermittency and multifractality of the time series has been verified as the prediction of the model. However, the ratios between wavelet coefficients describing the different hierarchical layers calculated backwards from the data have shown strongly negative correlation, while those are i.i.d. stochastic variables in usual multiplicative cascade models.

We have extended the multiplicative model to incorporate an additional stochastic term multiplied by the standard deviation of the variable. We have demonstrated through an empirical study and Monte Carlo simulations of the model that the proposed model is consistent with all the empirical results shown here.

It is noteworthy that the multiplicative cascade model and its extensions violate causality. Bacry, Delour and Muzy have proposed a stochastic process
along time axis designated as multifractal random walk keeping the essence of the multiplicative cascade model such as the multifractality and the correlations (Bacry et al. 2001). It might be accomplished in a forthcoming paper. 

\section*{acknowledgements}
This research was partially supported by a Grant-in-Aid for Scientific Research (C) No. 16K01259.




\section*{Appendix: WTMM method}
\label{sec:appendix}
This appendix briefly describes the WTMM method based on the continuous wavelet transform proposed earlier in the literature (Bacry et al. 1993; Muzy et al. 1993). The continuous wavelet transformation of the function $f$ using the analyzing wavelet $\psi$ is defined as
\begin{equation}
W_{\psi}[f](x,s)=\frac{1}{s}\int_{-\infty}^{\infty}f(u)\psi(\frac{u-x}{s})du,
\label{cwt}
\end{equation}
where parameters $s$ and $x$ respectively represent the dilation and the translation of the function $\psi$. 
The analyzing wavelet $\psi$ has been assumed to have $N_{\psi}>0$ vanishing moments. The successive derivative of the Gaussian function
\begin{equation}
\psi^{(N_\psi)}(x)=\frac{d^{N_\psi} (e^{-x^2/2})}{d x^{N_\psi}}
\label{aw}
\end{equation}
has $N_{\psi}>0$ vanishing moments. Here we specify $N_{\psi}=2$ and use the second derivative of the Gaussian function as the analyzing wavelet.
The WTMM method builds a partition function from the modulus maxima of the wavelet transform defined at each scale $s$ as the local maxima of $|W_{\psi}[f](x,s)|$ regarded as a function of x. Those maxima mutually connect across scales and form ridge lines designated as maxima lines.
The set $\mathcal{L}(s_0)$ is the set of all the maxima lines $l$ that satisfy
\begin{equation}
(x,s)\in l \Rightarrow s \le s_0,~\forall s \le s_0  \Rightarrow \exists(x,s)\in l.
\label{ml}
\end{equation}
The partition function is defined by the maxima lined as
\begin{equation}
Z(q,s)=\sum_{l\in \mathcal{L}(s)}(\sup_{(x,s')\in l}|W_{\psi}[x,s']|)^q.
\label{z}
\end{equation}
Assuming the power-law behavior of the partition function
\begin{equation}
Z(q,s)\sim s^{\tau(q)},
\label{tau}
\end{equation}
one can define the exponents $\tau(q)$.
The singular spectrum $D(\alpha)$ can be computed using the Legendre transform of $\tau(q)$:
\begin{equation}
D(\alpha)=\min_q (q\alpha -\tau(q)).
\label{D}
\end{equation}

\end{document}